\RequirePackage{lineno}

\documentclass[aps,prc,reprint,prcsuperscriptaddress,nofootinbib,11pt]{revtex4}  

\usepackage{epsfig}
\usepackage{rotating}
\usepackage{amssymb}
\usepackage{color}

\newcommand {\snn}  {\sqrt{s_{_{\rm NN}}}}
\newcommand {\gevc} {GeV/$c$}
\newcommand {\gevcc}    {GeV/$c^2$}
\newcommand {\chindf}   {\chi^2/{\rm ndf}}
\newcommand {\pip}  {\pi^{+}}
\newcommand {\pim}  {\pi^{-}}
\newcommand {\pipm} {\pi^{\pm}}
\newcommand {\Ks}   {K_S}

\newcommand {\Bvec} {\vec{B}}
\newcommand {\pt}   {p_{T}}
\newcommand {\minv} {m_{\rm inv}}
\newcommand {\mcut} {$\minv$$>$2~GeV/$c^2$}
\newcommand {\dgmcut}   {$\dg$(\mcut)}

\newcommand {\dg}   {\Delta\gamma}
\newcommand {\gOS}  {\gamma_{\rm OS}}
\newcommand {\gSS}  {\gamma_{\rm SS}}
\newcommand {\NOS}  {N_{\rm OS}}
\newcommand {\NSS}  {N_{\rm SS}}
\newcommand {\Nrho} {N_{\rho}}
\newcommand {\grho} {\gamma_{\rho}}
\newcommand {\dgCME}    {\dg_{\rm CME}}
\newcommand {\gCME} {\gamma_{\rm CME}}
\newcommand {\gMc}  {\gamma_{\rm m.c.}}

\newcommand {\mean}[1]  {\langle #1\rangle}

\begin{document}


\title{Isolating the chiral magnetic effect from backgrounds by pair invariant mass}
\author{Jie Zhao}
\email{zhao656@purdue.edu}
\affiliation{Department of Physics and Astronomy, Purdue University, West Lafayette, Indiana 47907, USA}
\author{Hanlin Li}
\affiliation{Department of Physics and Astronomy, Purdue University, West Lafayette, Indiana 47907, USA}
\affiliation{College of Science, Wuhan University of Science and Technology, Wuhan, Hubei 430065, China}
\author{Fuqiang Wang}
\email{fqwang@purdue.edu}
\affiliation{Department of Physics and Astronomy, Purdue University, West Lafayette, Indiana 47907, USA}
\affiliation{School of Science, Huzhou University, Huzhou, Zhejiang 313000, China}

\date{\today}

\begin{abstract}
Topological gluon configurations in quantum chromodynamics induce quark chirality imbalance in local domains, 
which can result in the chiral magnetic effect (CME)--an electric charge separation along a strong magnetic field. 
Experimental searches for the CME in relativistic heavy ion collisions via the charge-dependent azimuthal correlator ($\Delta\gamma$) 
suffer from large backgrounds arising from particle correlations (e.g.~due to resonance decays) coupled with the elliptic anisotropy. 
We propose differential measurements of the $\Delta\gamma$ as a function of the pair invariant mass ($\minv$), 
by restricting to high $\minv$ thus relatively background free, and by studying the $\minv$ dependence to separate the possible CME signal from backgrounds. 
We demonstrate by model studies the feasibility and effectiveness of such measurements for the CME search.
\end{abstract}




\keywords{Chiral magnetic effect, invariant mass, QGP, QCD}
\pacs{25.75.-q, 25.75.Gz, 25.75.Ld, 25.75.Ag}

\maketitle

\section{Introduction}
Vacuum fluctuations in quantum chromodynamics (QCD) can result in metastable local domains of gluon fields with non-vanishing topological charges~\cite{Lee:1974ma,Kharzeev:1998kz,Kharzeev:1999cz}. 
Interactions with those gluon fields can, under the approximate chiral symmetry restoration, change the overall chirality of quarks in those domains~\cite{Kharzeev:2007tn,Kharzeev:2007jp}. 
The chirality imbalance yields an electric charge separation under a strong magnetic field, a phenomenon called the chiral magnetic effect (CME)~\cite{Fukushima:2008xe,Muller:2010jd,Liu:2011ys}.
Strong magnetic fields are generated at early times by the spectator protons in relativistic heavy ion collisions, 
raising the possibility to detect the CME in those collisions~\cite{Voloshin:2004vk,Kharzeev:2004ey}. 
An observation of the CME would confirm a fundamental property of QCD and is therefore of great importance~\cite{Kharzeev:2015znc}. 
CME-like phenomena are not specific only to QCD and may have been observed in condense matter physics~\cite{Li:2014bha}.

A commonly used variable to measure the CME-induced charge separation in heavy ion collisions is the three-point correlator~\cite{Voloshin:2004vk}, 
\begin{equation}
\gamma\equiv\mean{\cos(\alpha+\beta-2\psi)}\,,
\end{equation}
where $\alpha$ and $\beta$ are the azimuthal angles of two particles and $\psi$ is that of the reaction plane (span by the beam and impact parameter directions of the colliding nuclei). 
Charge separation along the magnetic field ($\Bvec$), which is perpendicular to $\psi$ on average, 
would yield different values of $\gamma$ for particle pairs of same-sign (SS) and opposite-sign (OS) charges: $\gSS=-1, \gOS=+1$. 
However, there exist background correlations unrelated to the CME~\cite{Voloshin:2004vk,Wang:2009kd,Bzdak:2009fc,Liao:2010nv,Bzdak:2010fd,Schlichting:2010qia,Pratt:2010zn,Petersen:2010di,Toneev:2012zx}. 
For example, transverse momentum conservation induces correlations among particles enhancing back-to-back pairs~\cite{Bzdak:2009fc,Liao:2010nv,Bzdak:2010fd,Schlichting:2010qia,Pratt:2010zn}. 
This background is independent of particle charges, affecting SS and OS pairs equally and cancels in the difference, $\dg\equiv\gOS-\gSS$.
Recent experimental searches have thus focused on the $\dg$ observable~\cite{Kharzeev:2015znc,Zhao:2018ixy,Zhao:2018skm}; the CME would yield $\dg>0$. 
There are, however, also mundane physics that differ between SS and OS pairs. 
One such physics is resonance/cluster decays~\cite{Voloshin:2004vk,Wang:2009kd,Bzdak:2009fc,Liao:2010nv,Bzdak:2010fd,Schlichting:2010qia,Pratt:2010zn}, 
more significantly affecting OS pairs than SS pairs. 
Backgrounds arise from the coupling of elliptical anisotropy ($v_2$, a common phenomenon in heavy ion collisions~\cite{Heinz:2013th}) 
of resonances/clusters and the angular correlations between their decay daughters (nonflow)~\cite{Voloshin:2004vk,Wang:2009kd,Bzdak:2009fc,Schlichting:2010qia}.
Take $\rho\rightarrow\pip\pim$ as an example. 
The background is
$(\dg)_{\rho}=r_{\rho}\grho$, 
where $r_{\rho}=N_{\rho}/(N_{\pip}N_{\pim})$ is the relative abundance of $\rho$-decay pairs over all OS pairs, and 
$\grho\equiv\mean{f_{\rho}v_{2,\rho}}=\mean{\cos(\alpha+\beta-2\phi_{\rho})\cos2(\phi_{\rho}-\psi)}$
quantifies the $\rho$ decay angular correlations coupled with its $v_2$~\cite{Voloshin:2004vk,Wang:2016iov,Zhao:2018ixy,Zhao:2018skm}. 

Experimentally, significant positive $\dg$ values have been observed at the Relativistic Heavy Ion Collider (RHIC) 
and the Large Hadron Collider (LHC)~\cite{Abelev:2009ac,Abelev:2009ad,Abelev:2012pa,Adamczyk:2013hsi,Adamczyk:2014mzf}. 
The relative background and CME contributions are under extensive debate~\cite{QM17}. 
The recent observations of comparable $\dg$ in small system collisions~\cite{Khachatryan:2016got,Sirunyan:2017quh,Zhao:2017ckp,ZhaoJieQM17,Zhao:2017wck,Zhao:2018pnk}, 
where any CME signals would average to zero~\cite{Khachatryan:2016got,Belmont:2016oqp}, challenge the CME interpretation of the measured $\dg$ in heavy ion collisions. 
The major difficulty in distinguishing CME from backgrounds with the $\dg$ observable is their similar behaviors with respect to the event multiplicity~\cite{ZhaoJieQM17,Zhao:2017wck,Zhao:2018pnk}. 
This is because the magnetic field strength, to which the CME is sensitive, 
has a similar dependence as the $v_2$ (backgrounds) on the event multiplicity~\cite{Kharzeev:2004ey,Bzdak:2011yy,Deng:2012pc,Bloczynski:2012en,Chatterjee:2014sea}.
There have been various proposals and attempts to reduce or eliminate the backgrounds~\cite{Ajitanand:2010rc,Bzdak:2011np,Adamczyk:2013kcb,Wang:2016iov,Wen:2016zic,Acharya:2017fau,Sirunyan:2017quh}. 
The central idea is to ``hold'' the magnetic field fixed (in a narrow centrality) and vary the event-by-event $v_2$ from statistical 
and dynamical fluctuations~\cite{Voloshin:2010ut,Adamczyk:2013kcb,Chatterjee:2014sea}.
The first attempt was carried out by STAR~\cite{Adamczyk:2013kcb} where a charge asymmetry observable was analyzed as a function of the observed event-by-event $v_2$. 
A linear dependence was observed, expected from background, and the intercept was extracted representing a background-suppressed signal. 
ALICE~\cite{Acharya:2017fau} divided their data in each collision centrality according to $v_2$ in one phase space, 
and found the $\dg$ to be approximately proportional to the $v_2$ in the phase space of the $\dg$ measurement, consistent with background contributions. 
However, as recently pointed out by two of us~\cite{Wang:2016iov}, those methods suppressing background may not completely eliminate it. 
Another way to help search for the CME is to compare isobaric collisions~\cite{Voloshin:2010ut}, where the magnetic fields differ and the backgrounds are expected to be the same~\cite{Deng:2016knn}. 
However, these simple expectations may not be correct because of the non-identical isobaric nuclear structures~\cite{Xu:2017zcn}.

A new method to search for the CME, as we demonstrate in this article, 
is to eliminate the resonance background contributions using particle pair invariant mass ($\minv$) by (i) applying a lower cut on the $\minv$, 
(ii) fitting the low $\minv$ region by a two-component (TC) model, 
and (iii) fitting the low $\minv$ region by a two-component model aided by event-shape engineering (TC+ESE). 
We illustrate our method using the AMPT (A Multi-Phase Transport) model~\cite{Zhang:1999bd} and a toy {\em Monte Carlo} (MC) simulation. In both, the resonance masses are sampled from Breit-Wigner distributions~\cite{Agashe:2014kda,Lin:2014tya}. We use pions within pseudorapidity $|\eta|<1$ and $0.2<\pt<2$~\gevc, except otherwise specified.

It is worthwhile to note here, since the CME is generally a low $p_{T}$ phenomenon, that large $\minv$ region may not be the best place to search for it.
If no CME is found at large $\minv$, it does not necessarily mean that CME does not exist at low $\minv$. 
However, if a finite signal is observed at large $\minv$, where background sources diminish, 
it could be a strong evidence for the possible existence of the CME that would call for further investigations. 
It is also worthwhile to note that our low $\minv$ fitting procedure is not without assumptions. Although the background shape may be assessed by the ESE method, 
there is still the question of the CME shape as function of $\minv$. In principle, if the CME and the background have the same $\minv$ shape, 
then our invariant mass method would not be able to distinguish the two contributions. Theoretical inputs are much needed in this respect. 
Nevertheless, our method provides a potentially powerful way to extract the CME with reasonable assumptions of the $\minv$ dependence of the CME, 
which can be gauged to some extent by the fitting quality. 

The rest of the paper is organized as follows. 
Section \ref{sec:HM} presents the AMPT and toy-model studies by applying a lower $\minv$ cut to extract the $\dg$ in the high $\minv$ region. 
Section \ref{sec:TC} describes a two-component model fit to the $\dg(\minv)$. 
Section \ref{sec:ESE} extends the fitting method further by using ESE to determine the background $\dg(\minv)$ shape
to help extract the CME signal. 
Section \ref{sec:SM} puts our methods into experimental context and summarizes the paper.

\section{High-$\minv$ region}\label{sec:HM}
\subsection{A transport model study with null CME}
AMPT is a parton transport model~\cite{Zhang:1999bd}. It consists of a fluctuating initial condition, parton elastic scatterings, quark coalescence for hadronization, and hadronic interactions. 
The initial condition is taken from HIJING~\cite{Gyulassy:1994ew}. The string melting version~\cite{Lin:2001zk} is used in this study. 
Two-body elastic parton scatterings are treated with Zhang's Parton Cascade~\cite{Zhang:1997ej}, where the parton scattering cross section is set to 3~mb. 
After partons stop interacting, a simple quark coalescence model is applied to describe the hadronization process that converts partons into hadrons~\cite{Lin:2004en}. 
Subsequent interactions of these formed hadrons are modeled by a hadron cascade~\cite{Lin:2004en}. 
However, it is known that this version of the hadron cascade does not conserve charge, which is critical to the charge correlation study here. 
The hadronic scatterings, while responsible for the majority of the $v_2$ mass splitting, are unimportant for the main development of $v_2$~\cite{Li:2016flp,Li:2016ubw}, 
and thus may not be critical for the CME backgrounds. We thus turn off hadronic cascade in AMPT for our study here, as was done in Ref.~\cite{Ma:2011uma}. 
AMPT has been quite successful in describing variety of heavy ion data~\cite{Lin:2014tya}. 
It reproduces approximately the measured particle yields and distributions, and therefore should approximately describe those of resonances as well, 
which is relevant to the CME background study here. 
We simulate Au+Au collisions at $\snn=200$~GeV of various impact parameter ($b$) ranges.
For simplicity we use the known reaction plane in our analysis, fixed at $\psi=0$.

\begin{figure}[hbt]
  \begin{center}
    \includegraphics[width=0.6\textwidth]{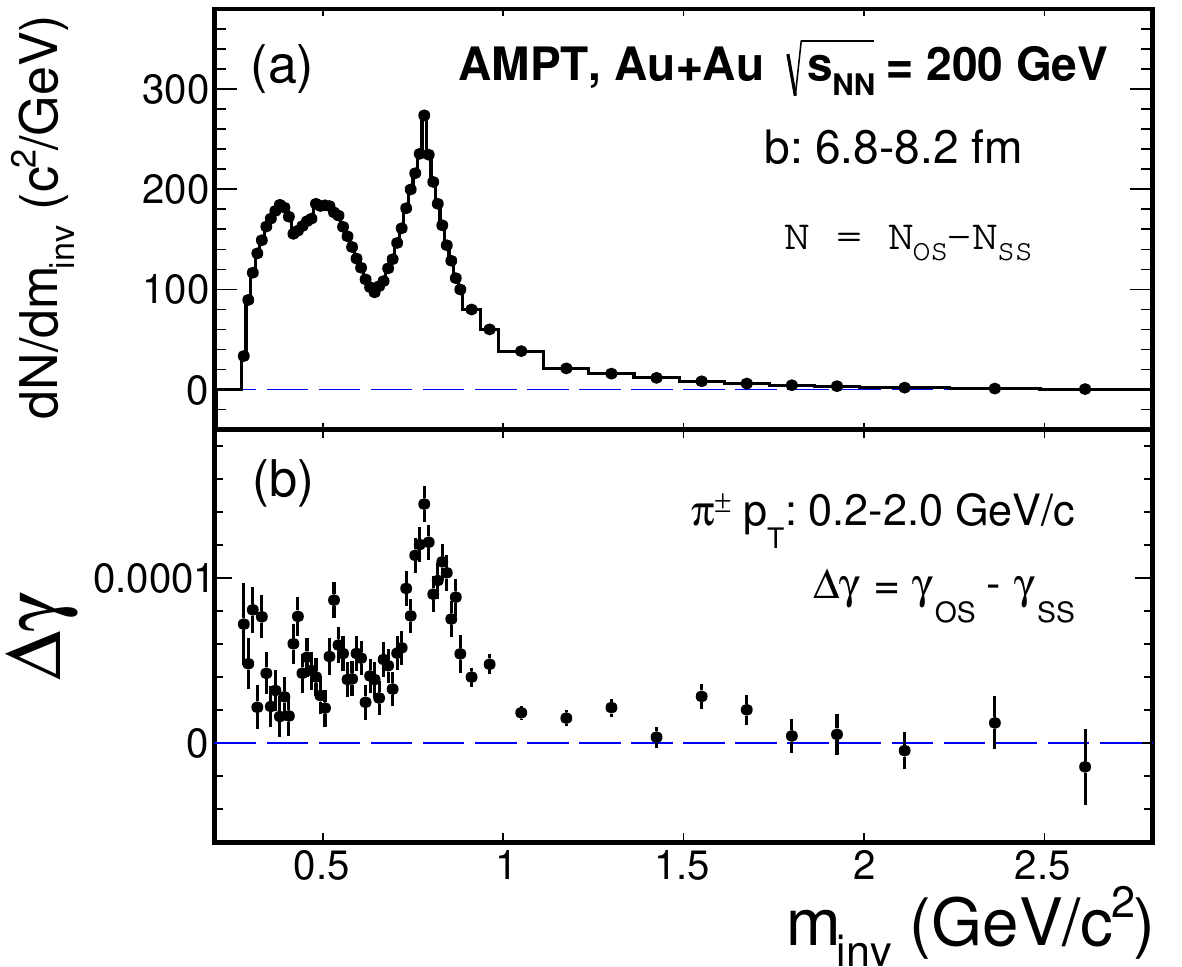}
    \caption{(a) Excess of opposite-sign (OS) over same-sign (SS) pion pairs, and (b) $\dg\equiv\gOS-\gSS$ as a function of pair invariant mass ($\minv$). Used are total $11\times10^6$ AMPT events of 200~GeV Au+Au collisions with $b=6.8$-8.2~fm.}
    \label{fig:ampt_m}
  \end{center}
\end{figure}

Figure~\ref{fig:ampt_m}(a) shows the $\minv$ distribution of the excess OS over SS pion pairs ($N\equiv\NOS-\NSS$), 
with $b=6.8$-8.2~fm (corresponding to the 20-30\% centrality of Au+Au collisions~\cite{Abelev:2008ab}, 
and average pion multiplicities $N_{\pip}\approx N_{\pim}\approx210$ within $|\eta|<1$). 
The $\rho$ peak is evident; the lower mass peaks are from Dalitz decays of $\eta$ and $\omega$ mesons (the $\Ks$ is kept stable in AMPT). 
Figure~\ref{fig:ampt_m}(b) shows the $\dg$ as a function of $\minv$. The $\rho$ contribution is clearly seen in the $\rho$ mass region. 
Since no CME is present in AMPT, the finite $\dg$ at $\minv\lesssim2$~\gevcc\ must be due to correlations from resonance decays, or generally, correlated pion pairs. 
This has been observed before~\cite{Ma:2011uma,ZhaoJieQM17}. For \mcut\ where resonance contribution to the OS over SS excess is small, the $\dg$ value is essentially zero, as expected.

\begin{figure}[hbt]
  \begin{center}
    \includegraphics[width=0.6\textwidth]{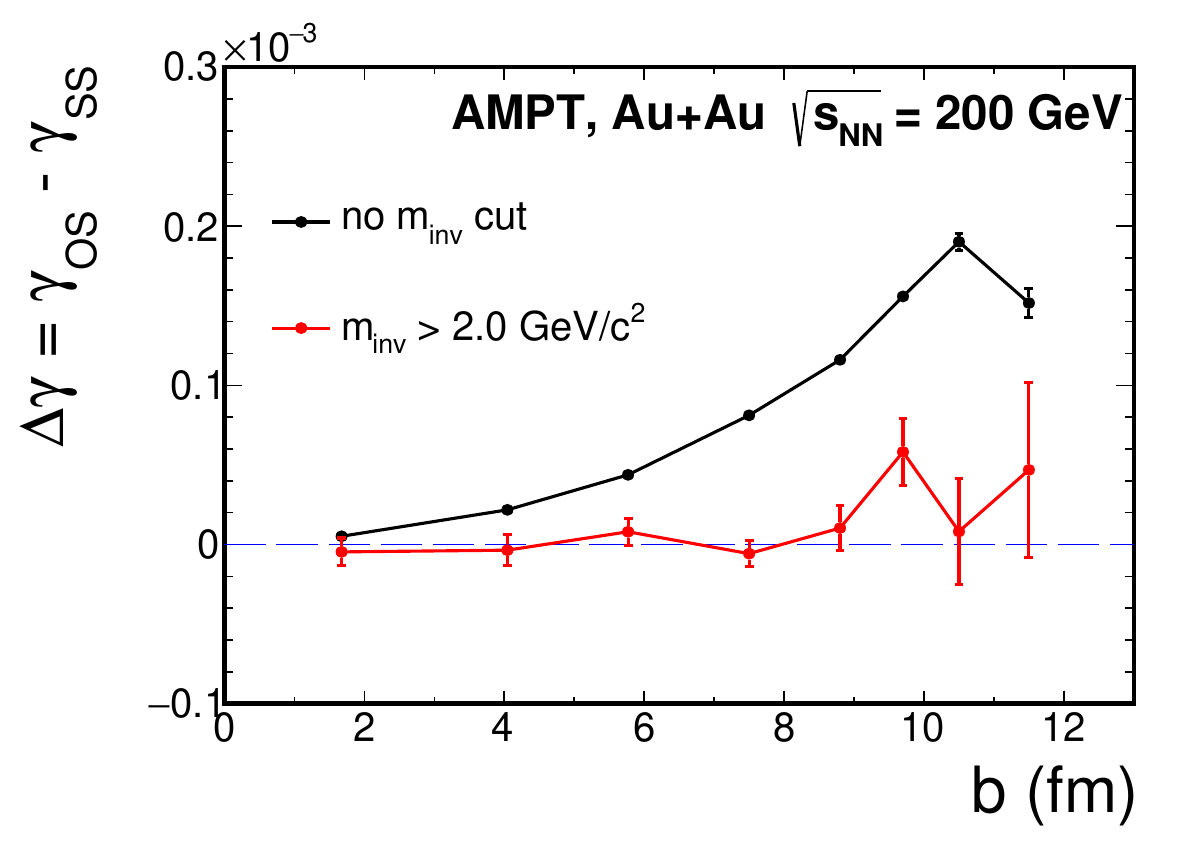}
    \caption{The $\dg$ as a function of impact parameter $b$ in 200~GeV Au+Au collisions by AMPT for all pion pairs (black markers) and for pairs with \mcut\ (red markers).}
    \label{fig:ampt_b}
  \end{center}
\end{figure}

Figure~\ref{fig:ampt_b} shows the $\dg$ in AMPT from all pairs and \dgmcut\ from pairs with \mcut. 
The positive $\dg$ is due to backgrounds; in \dgmcut\ this background is essentially eliminated, and as expected the result is consistent with zero. 
With the $11\times10^6$ AMPT events simulated for 200~GeV Au+Au collisions with $b=6.6$-8.2~fm, the inclusive $\dg$ value is $(8.1\pm0.1)\times10^{-5}$, and \dgmcut=$(-0.6\pm0.8)\times10^{-5}$. 
This represents a null signal with an upper limit of 20\% of the inclusive $\dg$ with 98\% confidence level (CL).

\subsection{A toy model study with finite CME}
In light of the AMPT results, we propose to apply a lower $\minv$ cut in real data analysis to search for the CME. 
We illustrate this point further by using a toy MC with input CME signal. 
Our toy model generates primordial $\pipm$, $\Ks$, and resonances ($\rho,\eta,\omega$), and decays the $\Ks$ and resonances (via both two- and three-body decays~\cite{Agashe:2014kda}). 
Particle kinematics are sampled according to
\begin{equation}
\frac{d^2N}{d\pt d\eta d\phi}=\frac{d^2N}{2\pi d\pt d\eta}\left(1+2v_2\cos2\phi+2a_1\sin\phi\right)\,,
\label{eq:toy}
\end{equation}
where $a_1$ is the CME signal parameter~\cite{Voloshin:2004vk}.
The particle $dN/dy$, $\pt$ spectra, and $v_2(\pt)$ correspond to the 40-50\% centrality of Au+Au collisions; 
they are as same as those used in Ref.~\cite{Adamczyk:2015lme} except that the primordial pion $\pt$ spectra are parameterized here with a better agreement with data at high $\pt$, 
and we have added $\Ks$. 
The pion multiplicities within $|\eta|<1$ are $N_{\pip}\approx N_{\pim}\approx100$ and those of primordial pions are $N^{\rm prim}_{\pip}\approx N^{\rm prim}_{\pim}\approx60$. 
The $\Ks$ multiplicity is taken to be 1/5 of the measured $\rho$'s~\cite{Adams:2003cc}, 
because some $\Ks$'s would have both their decay pions reconstructed as primary particles in experiments (such as STAR).
We generate $200\times10^6$ events with an input CME signal of overall strength 
$a_1=\pm0.008$ for primordial $\pipm$; for $\Ks$ and resonances $a_1=0$. Our input CME is independent of the particle $\pt$. 
This is supported by a recent theoretical study~\cite{Shi:2017cpu}, where the CME is insensitive to $\pt$ once $\pt$ is above 0.2~\gevc.

\begin{figure}[hbt]
  \begin{center}
    \includegraphics[width=0.6\textwidth]{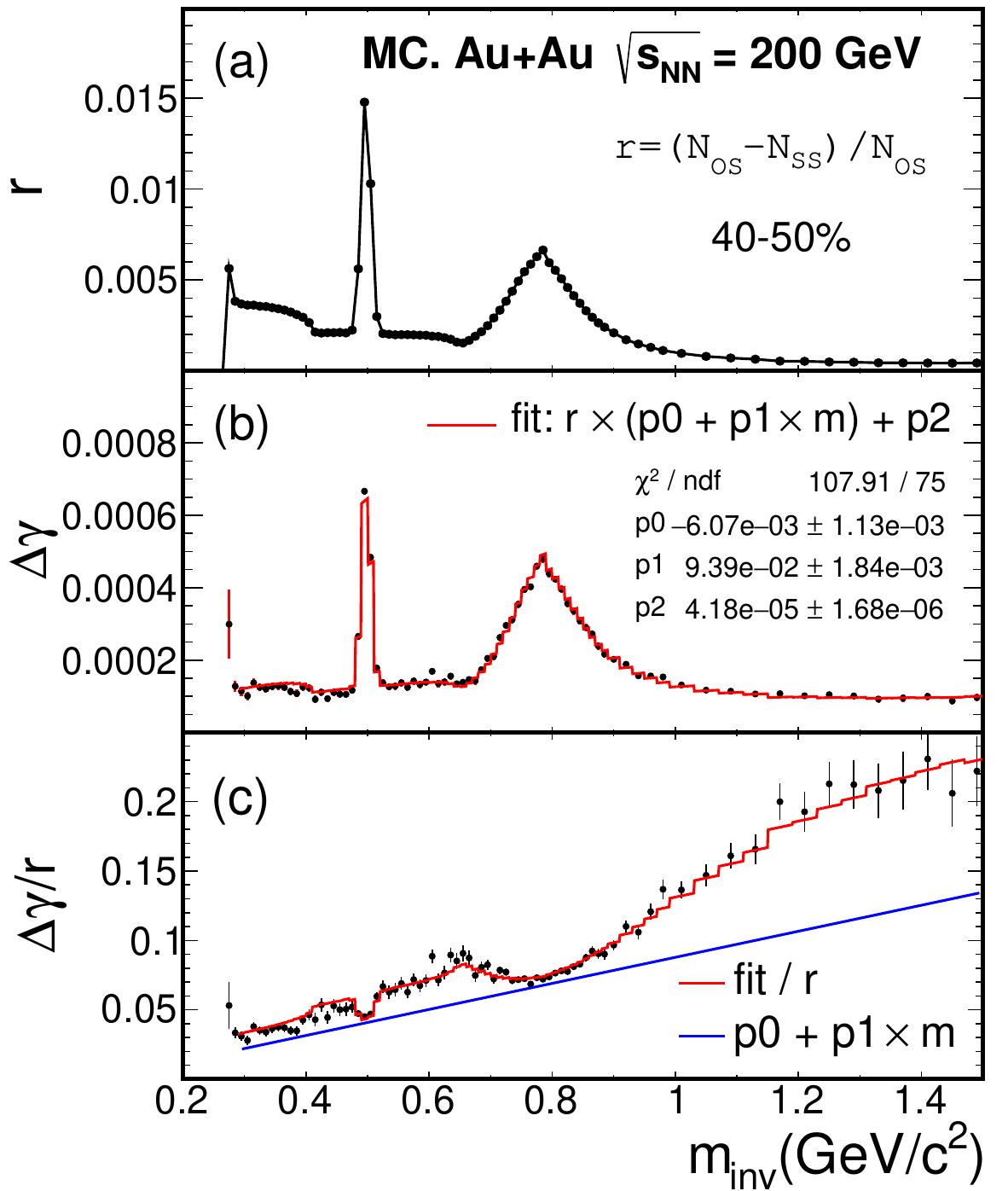}
    \caption{(a) $r=(\NOS-\NSS)/\NSS$, (b) $\dg$, and (c) $\dg/r$ as a function of $\minv$ from the toy MC simulation of total $200\times10^6$ events. Included are primordial $\pipm$, and $\Ks, \rho, \eta, \omega$ resonances, using parameters corresponding to measurements in Au+Au collisions of the 40-50\% centrality. A CME signal $a_1=\pm0.008$ is included.}
    \label{fig:toy}
  \end{center}
\end{figure}

Figure~\ref{fig:toy}(a) shows the relative OS pair excess, $r(\minv)\equiv(\NOS-\NSS)/\NOS$ as a function of $\minv$ from the toy MC. The $\Ks$ and $\rho$ peaks are evident. 
Figure~\ref{fig:toy}(b) shows the $\dg(\minv)$; the $\Ks$ and $\rho$ contributions are clear. 
The inclusive $\dg$ from Fig.~\ref{fig:toy}(b) is $(24.5\pm0.1)\times10^{-5}$; 
our input CME signal of $2a_1^2$, diluted by $(N^{\rm prim}_{\pi}/N_{\pi})^2$, is $4.6\times10^{-5}$, about 20\% of the inclusive $\dg$ value.
The $\dg(\minv)$ distribution in Fig.~\ref{fig:toy}(b) has a pedestal corresponding to the input CME signal. 
The pedestal extends to high $\minv$ (not shown) where resonance backgrounds vanish. 
A lower $\minv$ cut removes backgrounds to $r(\minv)$ but not the CME signal. 
The value \dgmcut\ $=(4.5\pm0.8)\times10^{-5}$ is consistent with the input CME signal, and it would present a $5\sigma$ measurement. 

The CME is generally believed to be a low-$\pt$ phenomenon~\cite{Kharzeev:2007jp}, and would thus be more prominent in the low $\minv$ region. 
With a $\minv>2$~\gevcc\ cut we used here, the particle average $\pt$ is typically 1.2~\gevc~\cite{WangQM18}. 
This is not very high and the CME may still be present above such a mass cut. 
Moreover, a recent study~\cite{Shi:2017cpu} indicates that the CME signal is rather independent of $\pt$ at $\pt>0.2$~\gevc, 
suggesting that the signal can persist to high $\minv$. 
Nevertheless, our proposal to apply a lower $\minv$ cut will eliminate resonance contributions to $\dg$; 
any measured remaining positive $\dg$ would point to the interesting possibility of the existence of the CME. 
A null measurement at high $\minv$, on the other hand, does not necessarily mean null CME also at low $\minv$. 

\section{Two-component model fit in low-$\minv$ region}\label{sec:TC}    
In what follows, we illustrate a fit method to potentially identify the possible CME at low $\minv$. 
Still use $\rho\to\pip\pim$ as an example, and consider the event to be composed of primordial pions containing CME signals ($\gCME$) and common (charge-independent) backgrounds, 
such as momentum conservation ($\gMc$)~\cite{Bzdak:2010fd,Pratt:2010zn}, and decay pions containing correlations from the decay~\cite{Voloshin:2004vk,Schlichting:2010qia,Wang:2016iov}. 
We have
\begin{equation}
  \dg=\frac{\NSS(\gCME+\gMc)+\Nrho\grho}{\NSS+\Nrho}-(-\gCME+\gMc)=r(\grho-\gMc)+(1-r/2)\dgCME\,.
  \label{eq:dg2}
\end{equation}
(If one normalized $\gOS$ by $\NSS$ instead, then Eq.~(\ref{eq:dg2}) would become simpler, 
$\dg'=r'\grho+\dgCME$ with $r'=(\NOS-\NSS)/\NSS$.)
Considering Eq.~(\ref{eq:dg2}), the $\dg$ can be expressed by two terms:
\begin{equation}
\dg(\minv)=r(\minv)R(\minv)+\dgCME(\minv)\,.
\label{eq:cme}
\end{equation}
The first term is resonance contributions, where the response function
\begin{equation}
R(\minv)\equiv\mean{f(\minv)v_2(\minv)}-\gMc
\end{equation}
contains $v_{2}$ of various resonances and is likely a smooth function of $\minv$ while $r(\minv)$ contains resonance spectral profile (Fig.~\ref{fig:toy}(a)). 
Consequently, the first term is not smooth but a peaked function of $\minv$.
The second term in Eq.~(\ref{eq:cme}) is the CME signal which should be a smooth function of $\minv$ (note we have dropped the negligible $r/2$). 
However, the exact functional form of $\dgCME(\minv)$ is presently unknown and needs theoretical input.
It is possible that the CME may possess some resonance shapes; in the extreme case where the CME has the same $\minv$ shape as the background, then our method would obviously fail.
Nevertheless, the $\minv$ dependences of the two terms are likely different, and this can be exploited to identify CME signals at low $\minv$. 
This is illustrated in Fig.~\ref{fig:toy}(c) where the ratio of $\dg/r$ is depicted. 
If CME signal is present, as is the case in our toy MC, $\dg/r$ should not be smooth, but with a deviation resembling the inverse shape of $r$ in Fig.~\ref{fig:toy}(a). 
This is clearly seen in Fig.~\ref{fig:toy}(c) in the $\rho$ mass region, although not as clear in the $\Ks$ mass region. 

In Eq.~(\ref{eq:cme}), $\dg(\minv)$ and $r(\minv)$ are measured, and $R(\minv)$ results from known physics and can in principle be obtained from models. 
AMPT indicates that $R(\minv)$ is a first-order polynomial. 
We can thus take a step further to fit the $\dg(\minv)$ in Fig.~\ref{fig:toy}(b) by Eq.~(\ref{eq:cme}) taking $R(\minv)$ as a first-order polynomial fit function, 
treating CME as a $\minv$-independent fit parameter (our input CME signal is $\pt$ independent). 
The fit result is superimposed as the red histogram in Fig.~\ref{fig:toy}(b), and in Fig.~\ref{fig:toy}(c) after divided by $r(\minv)$ from Fig.~\ref{fig:toy}(a). 
The straight line in blue in Fig.~\ref{fig:toy}(c) is the fit result for $R(\minv)$. 
The difference between the fit red histogram and the blue line is $\dgCME/r(\minv)$, which shows the inverse shape of $r(\minv)$. 
It is found, with the simulated statistics, that the inverse-shape feature becomes hard to identify when the CME input signal is smaller than 10\% of the inclusive $\dg$.
The fit parameters are written in Fig.~\ref{fig:toy}(b). The fit parameter for CME is $\dgCME=(4.2\pm0.2)\times10^{-5}$, not far away from the input CME signal of $4.6\times10^{-5}$. 
The fit $\chindf=108/75$ is not ideal because of the approximation for the $\minv$-dependence of $R(\minv)$, 
but it presents a potentially viable way to extract CME signals from data even at low $\minv$. 

Theoretically, the $\minv$ dependence of the CME is unknown. 
The likely sphaleron or instanton mechanism for transitions between QCD vacuum states~\cite{Kharzeev:1998kz,Kharzeev:2007jp,Fukushima:2008xe,Kharzeev:2004ey}, 
leading to the CME, might yield a broad $\minv$ distribution around $\minv\sim1$~\gevc. 
With such a CME signal, assuming a $\minv$-independent CME in our fit would probably still yield a reasonable average CME signal. 
To illustrate this point, we simulate sphalerons/instantons by a broad Gaussian mass distribution at 1~\gevcc\ with width 0.5~\gevcc. 
Each sphaleron/instanton is at rest and decays into a $\pip\pim$ pair where the decay polar angle is uniform in $\sin\theta$ 
and the azimuthal angles of $\pipm$ are sampled according to $dN/d\phi\propto1\pm\sin\phi$. 
The number of sphalerons/instantons is Poisson and on average 0.7\% (0.9\% within $|\eta|<1$) of the event single charge pion multiplicity. 
The pion azimuthal distributions in the form of Eq.~(\ref{eq:toy}) are $dN_{\pipm}/d\phi\propto1+2v_2\cos2\phi+0.009\times(1\pm\sin\phi)$.
This gives an effective CME signal of 
$2a_1^2\approx4.0\times10^{-5}$.
We apply our fit method, still assuming a constant CME, and obtain 
$\dgCME=(3.7\pm0.2)\times10^{-5}$,
consistent with the input signal, with $\chindf=187/75$. 

\section{Use ESE to determine background shape in two-component fit}\label{sec:ESE}
\begin{figure}
  \centering 
  \includegraphics[width=0.9\textwidth]{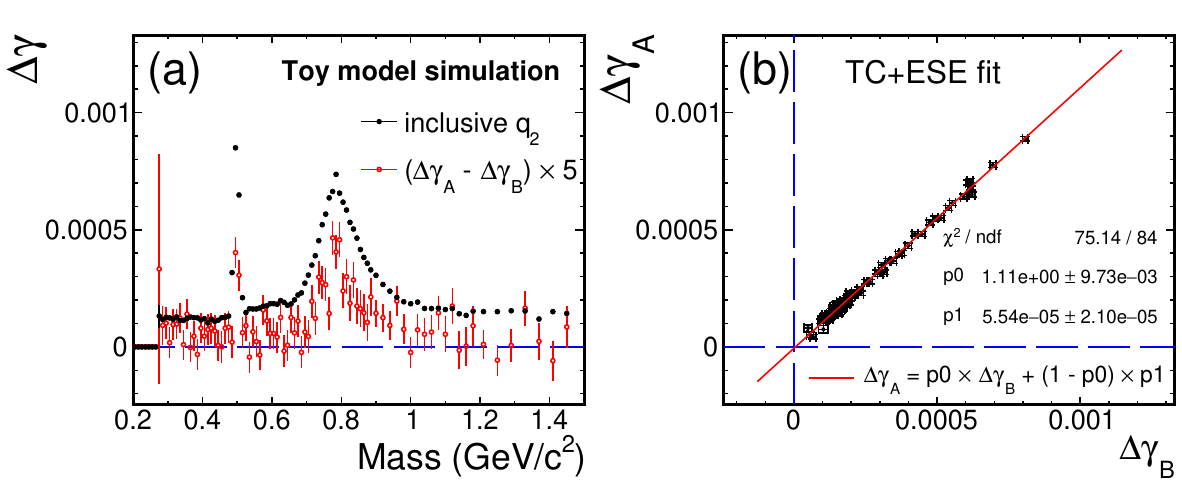}
  \caption{Toy model simulation of (a) inclusive $\dg$ and $\dg_A-\dg_B$ as functions of $\minv$, and (b) $\dg_A$ versus $\dg_B$. The simulation corresponds to the 40-50\% centrality Au+Au collisions at $\snn=200$~GeV. $\dg_A$ and $\dg_B$ are calculated in two event classes selected by $q_2$, corresponding to the 50\% high-$q_2$ and 50\% low-$q_2$ events, respectively. Each data point in panel (b) corresponds to one $\minv$ bin in panel (a).}
  \label{fig:ESE}
\end{figure}
In the TC fit in Sect.~\ref{sec:TC}, one needs the functional form of $R(\minv)$ as input. 
The linear form used in the Fig.~\ref{fig:toy} fit was motivated by AMPT results, but is strongly model dependent.
To lift this model dependency, one may resort to event-shape engineering (ESE)~\cite{Schukraft:2012ah,Acharya:2017fau,Sirunyan:2017quh} 
where events in each narrow centrality bin are divided into two classes according to the flow vector $q_2$:
\begin{equation}
  q_2 = \frac{1}{\sqrt{M}}\sum_{j=1}^{N}e^{in\phi_j}\,,
  \label{eq:q2}
\end{equation}
where $M$ is the particle multiplicity and $\phi_j$ is the $j$th particle azimuth.
The two $q_2$ event classes will have different average $v_2$ exploiting dynamical fluctuations of $v_2$~\cite{Schukraft:2012ah,Acharya:2017fau,Sirunyan:2017quh}. 
(In our simulation we included dynamical fluctuations of 40\% of the average $v_2$~\cite{Wang:2016iov}.) 
Since the magnetic fields are approximately equal for the two classes while the $v_2$ backgrounds differ, 
the difference in $\dg$ between the two event classes is a good representation of the background shape, and can therefore be used to add to the TC fit. 
We refer to such a fit as TC+ESE fit. Note that in this fit model the background is not required to be strictly proportional to $v_2$~\cite{WangQM18}.

We calculate $q_2$ using particles from $-1<\eta<-0.05$ and $\dg$ using particles from $0.05<\eta<1$, and vice versa (which we refer to as the subevent method). 
Figure~\ref{fig:ESE} left panel shows the $\minv$ distributions of $\dg_A-\dg_B$ (scaled by a factor of 5) together with the inclusive $\dg$. 
Note that, because of the different $\eta$ acceptances, the $\dg$'s in Fig.~\ref{fig:ESE} and Fig.~\ref{fig:toy} are numerically different.
The inclusive $\dg$ contains both background and CME. With the background shape given by $\dg_A-\dg_B$, the CME can be extracted from a fit $\dg=k(\dg_A-\dg_B)+\dgCME$. 
Since the same data are used in $\dg$ and $\dg_A-\dg_B$, their statistical errors are somewhat correlated. 
To properly handle statistical errors, one can simply fit the independent measurements of $\dg_A$ versus $\dg_B$, namely $\dg_A=b\dg_B+(1-b)\dgCME$ where $b$ and $\dgCME$ are the fit parameters. 
Figure~\ref{fig:ESE} right panel shows such a fit to our toy model simulation. 

\begin{figure}
  \centering 
  \includegraphics[width=0.6\textwidth]{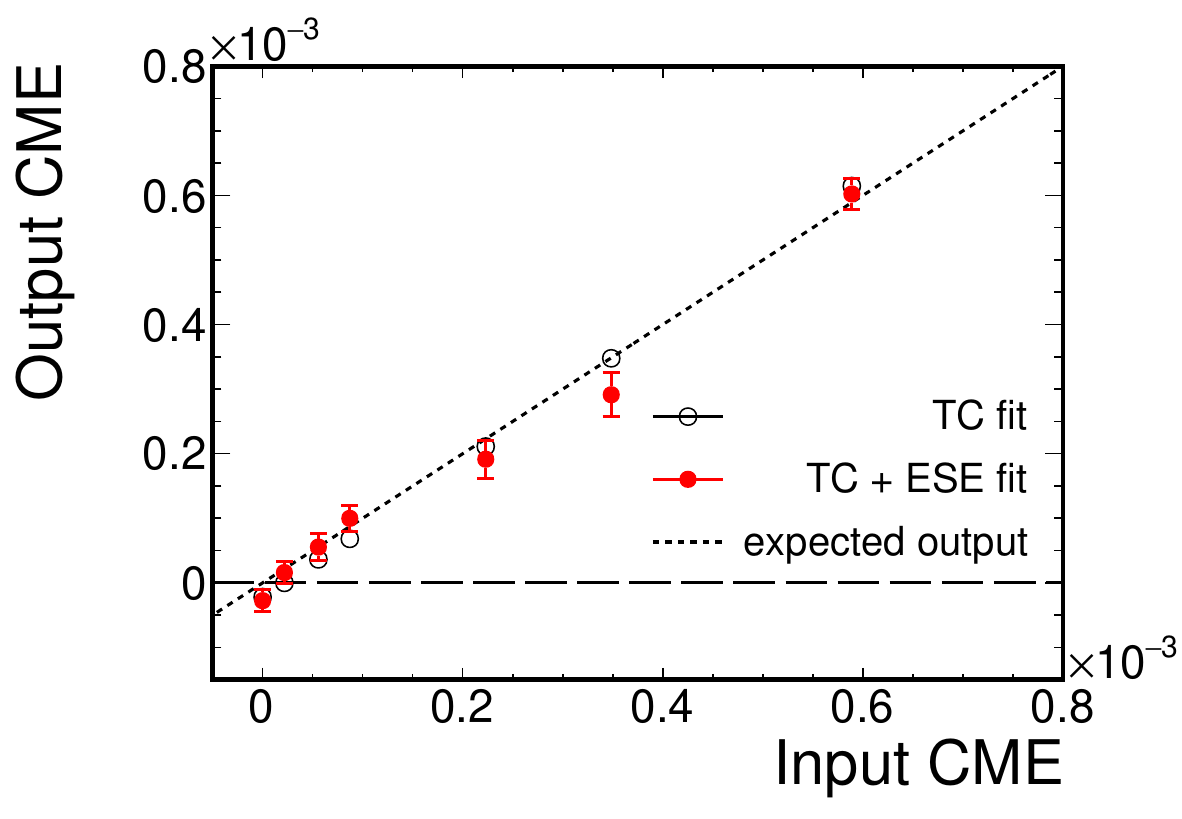}
  \caption{The $\dgCME$ parameter from the TC and TC+ESE fits versus the input CME signal.}
  \label{fig:CME}
\end{figure}

Figure~\ref{fig:CME} shows the $\dgCME$ parameters from the TC fit and the TC+ESE fit as functions of the input CME strength. 
The TC+ESE fit uncertainties on the $\dgCME$ parameter are relatively large. This is because $\dg_A-\dg_B$ is small (see Fig.~\ref{fig:ESE} left panel), 
corresponding to a relative difference in $v_2$ on the order of 10\% in our toy model simulation representing mid-central Au+Au collisions at RHIC. 
At the LHC, this difference is larger, approximately 20\%~\cite{Acharya:2017fau,Sirunyan:2017quh}, because of the larger event multiplicities than at RHIC. 
As seen from Fig.~\ref{fig:CME}, the fits appear to faithfully reproduce the input CME signal. 

\section{Discussion and summary}\label{sec:SM}
The STAR experiment at RHIC has accumulated Au+Au minimum bias data samples of $15\times10^6$ events from Run-4~\cite{Abelev:2009ac,Abelev:2009ad}, $57\times10^6$ events 
from Run-7~\cite{Adamczyk:2014mzf}, and $500\times10^6$ events from Run-11~\cite{Skokov:2016yrj}. 
If the CME signal is 1/3 of the measured inclusive $\dg$~\cite{Abelev:2009ac,Abelev:2009ad} and independent of $\minv$, these data samples (with the 20-60\% centrality) could yield, 
based on our toy MC study, a better than $5\sigma$ measurement of \dgmcut. If the CME signal is unobservable, then our analysis method could set, based on our AMPT result, an upper limit with 98\% CL on the CME of 5\% at \mcut\ relative to the measured inclusive $\dg$.
It is likely that the CME contribution decreases with $\minv$ and, depending on the detailed physics mechanism, may become difficult to observe at \mcut. 
Our fit methods can be used to explore and extract CME signals at low $\minv$. 
The methods rely on the rather robust assumption of different $\minv$ dependences of peaked resonance contributions and smooth CME signal, 
lifting the major difficulty of similar dependences of the background and CME on experimental variables thus far. 
The ESE method can further help determine the background dependence on $\minv$, thus improving the accuracy in the extraction of the CME signal.

In summary, topological charge fluctuations, resulting in the chiral magnetic effect (CME) and charge separation in relativistic heavy ion collisions, are fundamental properties of QCD. 
Experimental charge separation measurements by the azimuthal correlator ($\dg$) suffer from major backgrounds from resonance decays (generally, local charge conservation) coupled with elliptic anisotropy. 
In this article, we propose to measure the $\dg$ differentially as a function of the particle pair invariant mass ($\minv$). 
By using the AMPT (A Multi-Phase Transport) model, we demonstrate that one can essentially eliminate resonance decay backgrounds to $\dg$ by applying a lower cut on $\minv$. 
With a \mcut\ cut, an upper limit on the CME of 20\% of the inclusive $\dg$ can be achieved with $11\times10^6$ AMPT events of 200~GeV Au+Au collisions with impact parameter $b=6.6$-8.2~fm.
By using a toy {\em Monte Carlo} simulation with realistic resonance distributions and a $\pt$-independent input CME signal, we show that the resonance decay backgrounds are eliminated by the \mcut\ cut and the CME signal remains. 
With input CME signal of $a_1=0.008$ (20\% of the total $\dg$) and $200\times10^6$ events corresponding to the 40-50\% centrality of Au+Au collisions, a $5\sigma$ CME measurement could be achieved at \mcut.
We further show that one may be able to separate the presumably smooth CME signals from peaked resonance decay backgrounds in the low $\minv$ region by exploiting their different $\minv$ dependences. We show this by the toy MC with a $\pt$-independent CME signal as well as the signal from a broad sphaleron/instanton mass distribution.
We demonstrate that one may use the ESE to determine the background shape in $\minv$, thus help the extraction of the CME signal.

We note that our lower $\minv$ cut may also remove the CME signal together with the background. 
A negative conclusion at large $\minv$ does therefore not necessarily mean null CME also at low $\minv$. 
A positive signal at large $\minv$, however, would be a good indication of the possible CME. 
We also note that our two-component model fit in the low $\minv$ region is not without assumptions. 
One uncertainty is the unknown $\minv$ dependence of the CME, where theoretical guidance is much needed.
Nevertheless, our proposed invariant mass method provides a valuable tool and should help the ongoing experimental search for the CME at RHIC and the LHC. 

\section*{Acknowledgments}
We thank Prof.~Zi-Wei Lin for useful discussions. This work was supported in part by the National Natural Science Foundation of China Grant Nos.~11647306 and 11747312 and the U.S.~Department of Energy Grant No.~DE-SC0012910. HL acknowledges financial support from the China Scholarship Council.

\bibliographystyle{unsrt}
\bibliography{./ref}
\end{document}